# Comparison of the Mechanical Properties of Self-Assembled Langmuir Monolayers of Nanoparticles and Phospholipids

*Siheng Sean You[1], Rossen Rashkov[1], Pongsakorn Kanjanaboos[1,2], Ignacio Calderon,[1,3], Mati Meron[4], Heinrich Jaeger[1,2], Binhua Lin[1,4*]*

[1] James Franck Institute, The University of Chicago, Illinois 60637, USA.

[2] Department of Physics, The University of Chicago, Illinois 60637, USA.

[3] Departamento de Física, Universidad de Santiago de Chile, Santiago, Chile

[4] Center for Advanced Radiation Sources, The University of Chicago, Illinois 60637, USA.

**Abstract.** Nanoparticles with hydrophobic capping ligands and amphiphilic phospholipids are both found to self-assemble into monolayer films when deposited on the air/water interface. By separately measuring the anisotropic stress response of these films under uniaxial compression, we obtain both the 2D compressive and shear moduli of a range of different thin nanoparticle and phospholipid films. The compressive moduli of both nanoparticle and lipid films in the solid phase are on the same order of magnitude, whereas the shear moduli of the lipid films are found to be significantly lower. Additionally the moduli of the nanoparticle films depended substantially on the polydispersity of the constituent particles – broader size distribution lowered the stiffness of the nanoparticle film.

**Introduction**. Amphiphilic surfactants spontaneously self-assemble into Langmuir monolayers when deposited on the surface of an aqueous subphase. The most studied class of these materials is of biological origin: amphiphilic proteins and lipids. These monolayers allow for exploration of the rheological properties of biological membranes. Phospholipids have been used as a model to study the mechanical properties of cell membranes and pulmonary surfactants.[1, 2] Recently, there has also been a developing interest in a different class of Langmuir monolayers, composed of inorganic nanoparticles capped with organic surfactants. Like amphiphilic molecules, these capped nanoparticles self-assemble into 2D lattices at the air-water interface.[3-14] Nanoparticles exhibit remarkable physical properties including quantum confinement,[15] plasmon resonance[9, 16-19] and superparamagnetism[5, 20] Thin films composed of nanoparticles have shown potential applications in biosensors[16, 19, 21], high sensitivity resonators[14], filtration devices[22], magnetoresistive devices[10], and flexible electronics.[23-25]

Langmuir monolayers in particular have remarkable mechanical properties as two-dimensional materials that can extend into the third dimension. Parallels can be drawn between nanoparticle and lipid films. Both of these films exhibit fascinating elastic responses: upon compression these systems show a rich morphology of folding and buckling phenomena. Monolayers assembled from mixtures of lipids can, upon lateral compression, collapse into the third dimension either through folding or vesicle formation.[1, 2, 26, 27] Nanoparticle films, upon compression, undergo transitions from monolayers to multilayers and can buckle, wrinkle or fold out of plane.[3, 4, 8] These phenomena arise out of a complex series of interactions between the constituent materials of Langmuir films. The head and tail groups of lipids determine the intermolecular interactions and the packing behavior of their self-assembled membranes.[28] Similarly, nanoparticles have particle-particle interactions defined by factors including inorganic core attraction, ligand-ligand repulsion/van der Waals interactions, and entropic considerations of ligand packing.[29-32] These properties not only influence the ordering of nanoparticles but also dictate the mechanical response and subsequent collapse behavior of assembled monolayers. A quantitative investigation of the mechanical properties of these two classes of films can help us gain insight into similarities and differences in their collapse behaviors.

Previous efforts have focused on the measurement of elastic properties of freestanding nanoparticle films. Through use of tapping AFM, the Young's moduli of these membranes have been measured.[7, 12] The Poisson ratio of an Au nanoparticle membrane has also been directly measured by SEM while inducing strain on an Au nanoparticle film.[13] Together these methods can provide a complete picture of the elastic properties of freestanding nanoparticle membranes. Alternatively, in supported films that exhibit wrinkling behavior, the bending rigidity of the membrane has been determined through measuring the wavelength of the buckling patterns on the surface of the film.[8]

Methods of measuring rheology of Langmuir monolayers of organic molecules have been well developed. Of particular interest, Petrov et al. proposed measuring the anisotropic stress response of a protein film under continuous uniaxial compression as a method to obtain the compressive and shear moduli of their monolayers.[33] This method has been previously extended to measure the elastic properties of supported silica nanoparticle monolayers.[34]

Measurements of anisotropic stress response offer several advantages that can be used to expand our understanding of the mechanical properties of supported nanoparticle and lipid membranes. Firstly, this method allows us to extract two sets of mechanical parameters, the 2D compressive and shear moduli. Previously, two separate experiments were necessary to determine each elastic parameter. Secondly, it explores the mechanical properties of metal/metal oxide nanoparticle monolayers on a macroscopic length scale – the self-assembled monolayers studied in these experiments have dimensions on the order of 10-100 cm$^2$, whereas earlier work on the mechanical properties of nanoparticle monolayers focused on freestanding membranes with sizes on the order of 10-100 $\mu$m$^2$. Finally, this method, being able to measure both lipid and nanoparticle films, allows for the first direct, quantitative, comparison of the mechanical properties of these films.

In this paper we report measurements of the 2D compressive and shear moduli of a series of different nanoparticle and lipid monolayers by measuring their anisotropic stress responses under continuous uniaxial compression. Films of 1,2-dipalmitoyl-*sn*-glycero-3-phosphocholine (DPPC) and 1-palmitoyl-2-oleoyl-sn-glycero-3-phospho-(1'-rac-glycerol) (POPG), well-studied lipid constituents of pulmonary surfactant are investigated. The mechanical properties of Au, Ag, and Fe$_3$O$_4$ nanoparticle films are measured, and their qualitative mechanical responses are also characterized through optical microscopy of the film during compression.

**Experimental.**

**A. Sample preparation and film deposition** Different nanoparticle samples were purchased from Ocean Nanotech: 5.1 $\pm$ 1.2 nm Au (Lot 092211) ligated with dodecanethiol, 3.6 $\pm$ 1.5 nm Ag ligated with dodecanethiol (Lot B106B), and 13.8 $\pm$ 0.4 nm Fe$_3$O$_4$ nanoparticles ligated with oleic acid (Lot 031910). TEM pictures of the samples are shown in Figure 1. Additionally, size-controlled Au-nanoparticles (4.8 $\pm$ 0.9 nm) were obtained through the selection procedure mentioned below. The Fe$_3$O$_4$ and Ag samples were dispersed in chloroform to a concentration of 0.5 mg/ml, and the Au-nanoparticle samples were dispersed in toluene at a concentration of 1 mg/ml. TEM of Au and Fe$_3$O$_4$ nanoparticles were taken after drop-casting the nanoparticle solution onto a TEM grid. The TEM images of the Ag nanoparticles were taken by Ocean Nanotech. DPPC and POPG samples were purchased from Avanti Polar Lipids, and dispersed in chloroform to a concentration of 1 mg/ml. Langmuir films were formed by the deposition of solution onto the water/air interface of a NIMA 601 Langmuir trough. A droplet of nanoparticle solution was allowed to form on the tip of a needle and was touched to the surface of the air/water interface on a Langmuir trough. This process was repeated until a film covering the surface of the Langmuir trough was assembled. After film deposition, the sample was left to equilibrate for fifteen minutes prior to compression.

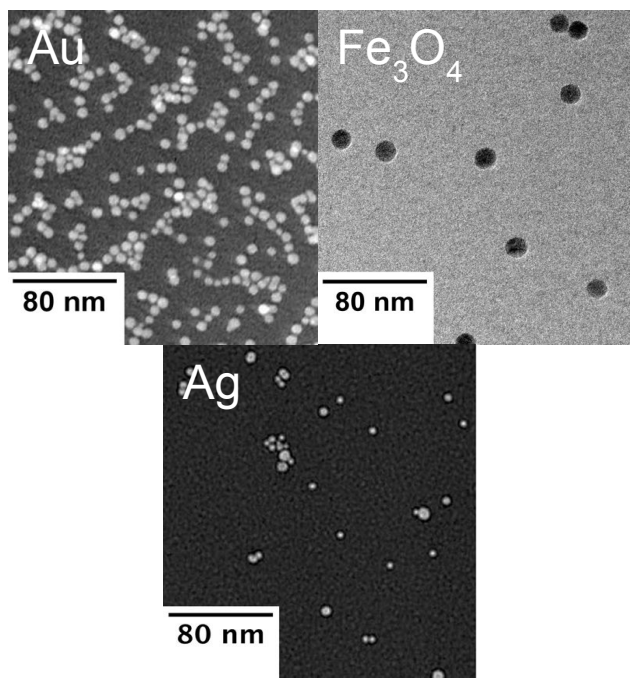

**Figure 1.** TEM Images of Au nanoparticle (4.8 ± 0.9 nm for size selected sample, 5.1 ± 1.2 nm for non-size selected sample), Fe$_3$O$_4$ nanoparticles (13.8 ± 0.4 nm) and Ag nanoparticles (3.6 ± 1.5 nm).

**B. Nanoparticle Size Selection** Ethanol was added to Au-nanoparticles suspended in toluene until the solution became opaque. The solution was then centrifuged, and the supernatant solution was recovered. Additional ethanol was added to the recovered supernatant and the solution was subsequently centrifuged again. The precipitate was recovered and suspended in toluene. The size distribution of the particles were measured using TEM, and the differences in the size distributions between the non-size selected and size selected particles are shown in Figure 2.

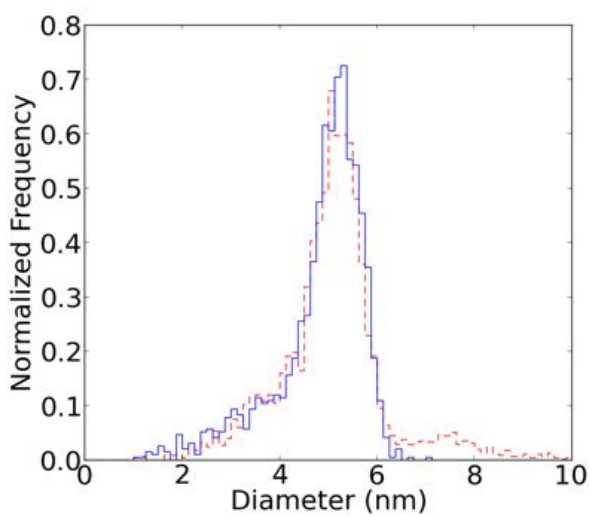

**Figure 2.** Histogram of size distributions of Size Selected (Blue) and Non Size Selected (Red) Au-nanoparticles.

**C. Langmuir Trough and surface pressure measurements**. Our NIMA Langmuir trough (Fig. 3) had a fully opened area of ~80 cm$^2$. The width of the trough was 7 cm, and the length of the trough was allowed to vary to control the area of the trough. Before each experiment the Langmuir trough was cleaned 3 times with chloroform. After cleaning, the barriers of the trough were compressed, and the initial surface pressure of pure water was found to be < 0.1 mN/m at maximal compression. Two NIMA pressure sensors with attached Wilhemy plates were oriented perpendicular and parallel to the confining barriers, and placed in the center of the Langmuir trough. Surface pressure measurements were taken throughout the duration of the compression. Our experiments were conducted at room temperature, around 22° C. The trough area was reduced at a rate of 5 cm²/min, unless otherwise noted, corresponding to a linear speed of 0.71 cm/min.

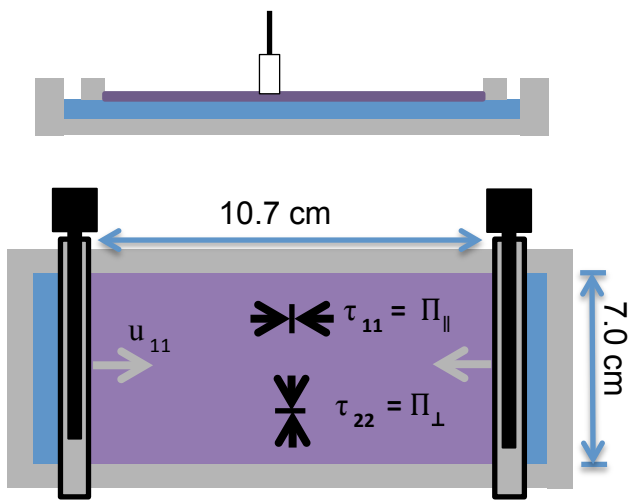

**Figure 3. Top**: Side view of Langmuir trough with Wilhelmy plate. **Bottom**: A top-down view of Langmuir trough, showing positions of Wilhelmy plates parallel and perpendicular (relative to the movable barriers) used to measure stresses ($\tau_{11}$ and $\tau_{22}$ respectively) induced by uniaxial compression ($u_{11}$) by movable barriers.

**D. Optical Microscopy.** The experimental set up used was similar to the one used by Leahy et al.[8] The nanoparticle films were imaged using an Olympus BH-3 optical microscope equipped with a color Leica CCD camera connected to a computer and recorded during compression using Streampix 3.0 software. Images were captured at a 50x magnification. One pressure sensor was used to concurrently take surface pressure measurements to correlate the optical microscopy with the 2 pressure sensor data.

**Results and Discussion**

**Theoretical Background.** The simplest model of elasticity is Hooke's Law: $\tau = E\mu$. The stress ($\tau$), or force experienced by a material is directly proportional to the strain of the system ($\mu$) or its deflection from equilibrium, multiplied by a constant. To account for possible dissipation effects in the material, a viscous damper can be placed in series with the spring. This results in the well-known Maxwell model in one dimension; in this case the stress ($\tau$) on the system can be described as a function of the viscosity ($\zeta$) and elastic modulus or spring constant ($E$):

$$\tau = E\mu + \zeta D \quad (1)$$

This model is able to describe materials that exhibit a response dependent both on the strain ($\mu$) and the rate of strain ($D$).

An analogous expression exists in two dimensions:

$$\tau_{11} = (E+G)\mu_{11} + (E-G)\mu_{22} + (\zeta+\eta)D_{11} + (\zeta-\eta)D_{22} \quad (2)$$

$$\tau_{22} = (E-G)\mu_{11} + (E+G)\mu_{22} + (\zeta-\eta)D_{11} + (\zeta+\eta)D_{22} \quad (3)$$

where $\tau$, $\mu$, and $D$ are the stress, strain, and rate-of-strain components, respectively. In two dimensions, two elastic and viscous moduli are necessary to characterize the system: $E$ and $G$ are the compressive and shear elastic moduli, while $\zeta$ and $\eta$ are the compressive and shear viscosities. Equations (2) and (3) can be simplified by imposing several conditions. If deformation of the film only occurs in a single direction, such as during uniaxial compression in the experimental setup shown in Figure 1, then $\mu_{22} = 0$ and $D_{22} = 0$. Additionally, if the film primarily exhibits elastic response, then the rate of strain dependent viscosity terms are negligible ($\zeta = \eta = 0$). This claim was tested for the nanoparticle films by examining the isotherms of 15 nm $Fe_3O_4$ films compressed at different barrier speeds (5, 10, 20 cm²/min). We found that the pressure response of the systems (Fig. 2) were within experimental error, for the monolayer region of the isotherm. This shows that viscosity is negligible for nanoparticle films in our region of interest.

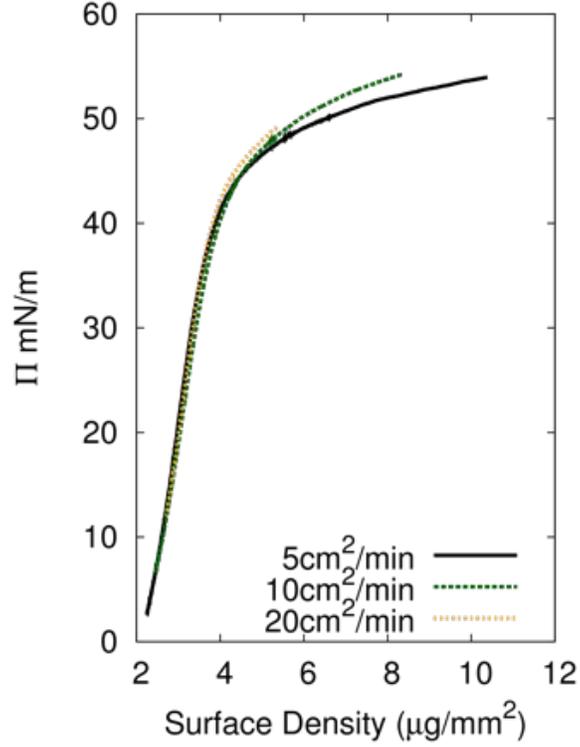

**Figure 4.** Pressure vs Surface Concentration Isotherms for 15nm $Fe_3O_4$ nanoparticle film taken at different (5,10,20 $cm^2$/min) compression speeds. Varying compression speeds does not have a significant effect on the shape of the isotherm in the monolayer regime between 0 and ~40 mN/m, where the monolayer moduli are measured.

Using these simplifications, the equations become

$$\tau_{11} = (E + G)\mu_{11} \quad (4)$$

$$\tau_{22} = (E - G)\mu_{11} \quad (5)$$

Assuming that the compression is quasistatic ($\tau_{11} = \delta\Pi_\parallel$, $\tau_{22} = \delta\Pi_\perp$) and using the Cauchy definition of strain, $\mu_{11} = \frac{\delta A}{A}$, the above expressions can be simplified into the following equations:

$$-A \frac{\delta \Pi_\parallel}{\delta A} = (E + G) \quad (6)$$

$$-A \frac{\delta \Pi_\perp}{\delta A} = (E - G) \quad (7)$$

Solving for E and G, we obtain the final expressions as:

$$E = -\frac{A}{2}\left(\frac{\delta \Pi_\parallel}{\delta A} + \frac{\delta \Pi_\perp}{\delta A}\right) \quad (8)$$

$$G = -\frac{A}{2}\left(\frac{\delta \Pi_\parallel}{\delta A} - \frac{\delta \Pi_\perp}{\delta A}\right) \quad (9)$$

**Method for extracting E and G.** Our experiment measures a surface pressure ($\Pi_\parallel$ and $\Pi_\perp$) vs area (A) isotherm. By taking the numerical derivative of the isotherm, we obtain $\frac{\delta \Pi_\parallel}{\delta A}$ and $\frac{\delta \Pi_\perp}{\delta A}$ and are able to determine the $E$ and $G$ of our Langmuir monolayers. The $\Pi - A$ isotherms are empirically fit to the sum of a 4th degree polynomial and an error function. This fitting function has no physical significance; it is used to smooth the isotherms so that the numerical derivatives can be extracted. To normalize for trough area, the isotherms are plotted in terms of $\Pi$ vs surface mass density of sample.

Previous X-ray and TEM experiments[3-6] indicate the pressure regimes in which the different nanoparticle films remain in the monolayer phase. We focus on comparing the compressive and shear moduli of our nanoparticle and lipid films in these regions, which are indicative of monolayer Langmuir films.

**Lipid Films.** The measured DPPC isotherm (Fig. 3) agrees well with those in the literature. DPPC at large area per molecule remains in the 2D gaseous phase. Upon compression, the lipid first enters a liquid phase and then undergoes a phase transition with two-phase coexistence into the solid phase. The measured compressive moduli in the liquid phase agree with those observed previously using other methods, between 20 and 30 mN/m[35], and the measured DPPC compressive modulus (~200 mN/m) in the solid phase preceding the collapse is similar to that reported in previous papers.[36, 37] At surface pressures greater than 50 mN/m, our measured compressive moduli in the solid phase of DPPC becomes lower than that given in previous literature values as the Langmuir trough used in these experiments maintains a positive meniscus, making high surface pressures impractical for the experimental apparatus. The shear modulus of the DPPC remains zero in the gaseous and liquid phases, as expected. However, even following the phase transition into the solid phase, the shear modulus remains relatively low (1-5 mN/m). The "solid" phase of DPPC has been better qualified as a type of crystalline phase consisting of small crystal domains with different tail group orientations.[2, 28] This powder-like structure would have a reduced shear modulus compared to that of a conventional solid. The shear modulus of DPPC measured using a rheometer has previously been shown to have a frequency dependence, but the reported values[37] (0.1-10 mN/m) are close to those measured here.

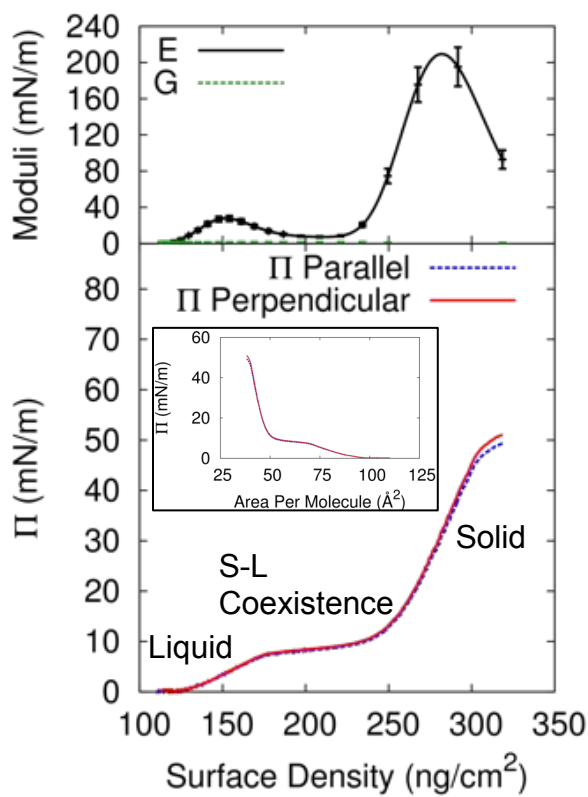

**Figure 5.** (Bottom) DPPC Surface Pressure/Surface Density and (Top) 2D Moduli/Surface Density. The standard phases, (gas, liquid, solid-liquid (S-L) coexistence, solid) are labeled. **Inset:** DPPC isotherm in terms of Surface Pressure vs Area Per Molecule, typically used when studying DPPC monolayers.

The POPG film, as see in Figure 6, has a low compressive modulus between 10 and 30 mN/m and zero shear modulus. As POPG remains in the liquid phase during room temperature, its vanishing shear modulus is to be expected. Unlike DPPC, POPG has an unsaturated hydrocarbon tail that introduces disorder into the packing of the lipid. As it remains in the liquid phase, the compressive modulus of the pure POPG film remains relatively low, comparable to the compressive modulus of DPPC in the liquid phase, but substantially lower than the compressive modulus of DPPC in the solid phase.

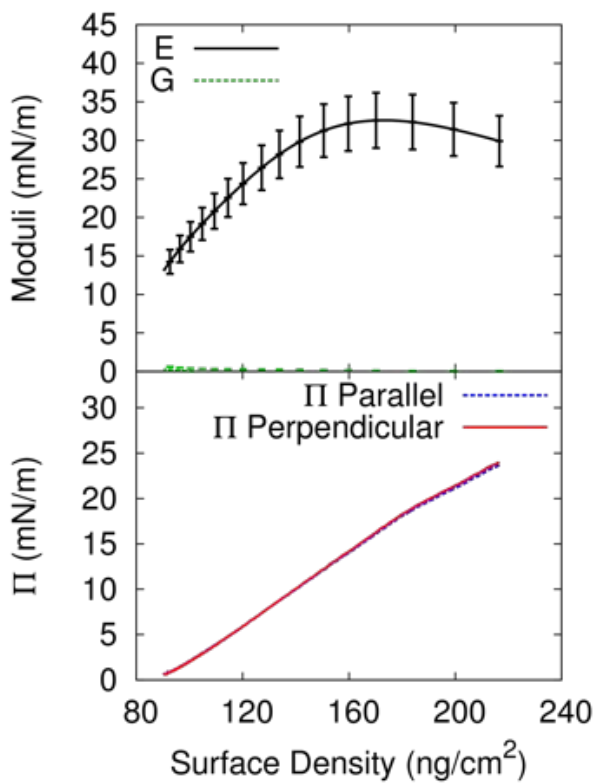

**Figure 6.** (Bottom) POPG Pressure/Density Isotherm and (Top) 2D Moduli/Density. As room temperature is greater than the melting temperature of POPG, the liquid remains in the liquid phase and, therefore, does not undergo any phase transitions.

**Table 1.** Lipid Monolayer Peak Modulus

| **Lipid Monolayer** | **Peak E (mN/m)** | **Peak G (mN/m)** |
|---|---|---|
| **DPPC** | 198 $\pm$ 7 | 4 $\pm$ 2 |
| **POPG** | 32 $\pm$ 8 | 0 $\pm$ 0.5 |

**Nanoparticle Films** The isotherms (Figures 7,8) and optical images (Figure 9) indicate that the various nanoparticle systems undergo a similar series of morphological transitions upon compression, differing mainly in their mode of collapse. At the start of the barrier compression, each nanoparticle sample consists of numerous particle rafts separated by void space of open air/water interface. Consequently, the measured macroscopic compressive and shear moduli are relatively low.

**Table 2.** Nanoparticle Monolayer Peak Modulus

| Nanoparticle Monolayer | Peak E (mN/m) | Peak G (mN/m) | Nanoparticle Size (Diameter, nm) |
|---|---|---|---|
| **15nm $Fe_3O_4$** | 58 $\pm$ 5 | 15 $\pm$ 1 | 13.9 $\pm$ 0.4 |
| **3 nm Ag** | 85 $\pm$ 35 | 17 $\pm$ 10 | 3.6 $\pm$ 1.4 |
| **5 nm Au** | 180 $\pm$ 7 | 34 $\pm$ 3 | 5.1 $\pm$ 1.2 |
| **5 nm Au (size selected)** | 450 $\pm$ 9 | 63 $\pm$ 17 | 4.8 $\pm$ 0.8 |

Upon further compression, these rafts begin to merge, causing a steep increase in surface pressure and a corresponding increase in shear and compressive moduli. Eventually, the film becomes one optically uniform monolayer, at which point the surface pressure begins to rise sharply. In this region, previous X-ray scattering measurements for Ag and Au-nanoparticles, and TEM measurements for $Fe_3O_4$ nanoparticles show that films remain in the monolayer regime[3-5]. Both the compressive and shear moduli attained their maximal valves (Table 2) near the end of the monolayer regime, which should correspond to the stiffness of a macroscopic, supported nanoparticle monolayer before it undergoes collapse into the third dimension.

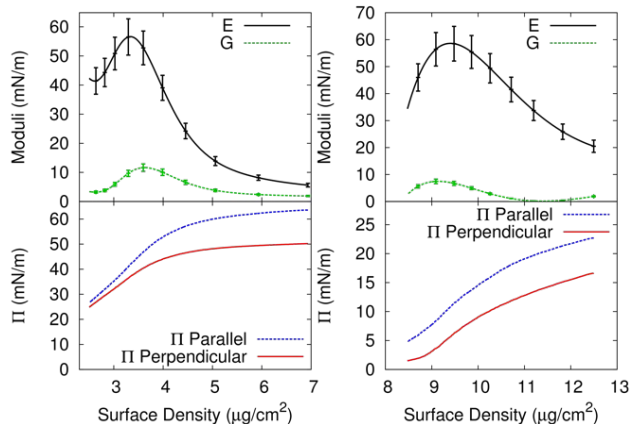

**Figure 7. Left:** Pressure/Density Isotherms (Bottom) and 2D Moduli/Surface Density (Top) of 13.8 nm $Fe_3O_4$-nanoparticle film. **Right:** Pressure/Density Isotherms (Bottom) and 2D Moduli/Surface Density (Top) of 3.6 nm Ag-nanoparticle film

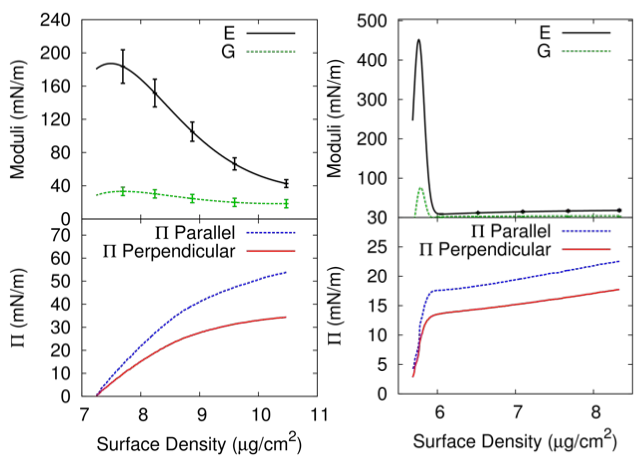

**Figure 8.** Pressure/Density Isotherms (Bottom) and 2D Moduli/Surface Density (Top) of non-size selected 5.1 nm Au-nanoparticle film. **Right:** Pressure/Density Isotherms (Bottom) and 2D Moduli/Surface Density (Top) of size selected 4.8 nm Au-nanoparticle film

Further compression of the film past the monolayer regime results in the formation of folding patterns, which correspond to a decrease in the slope of the isotherm and a corresponding decrease in compressive and shear moduli. For the Au-nanoparticles, dark lines appear on the film (Fig. 9a), which is a signature of a monolayer-to-trilayer transition.[8] Ultimately, the film buckles out of the plane and then collapses irreversibly by forming localized folds (Fig. 9b). The Ag nanoparticles show a transition to a multilayer, previously characterized as a monolayer-bilayer[4] followed by similar buckling and folding patterns (Fig. 9d). The 15 nm $Fe_3O_4$ particles formed a mixture of localized smaller and larger folds (Fig. 9c), running preferentially parallel to the barriers.

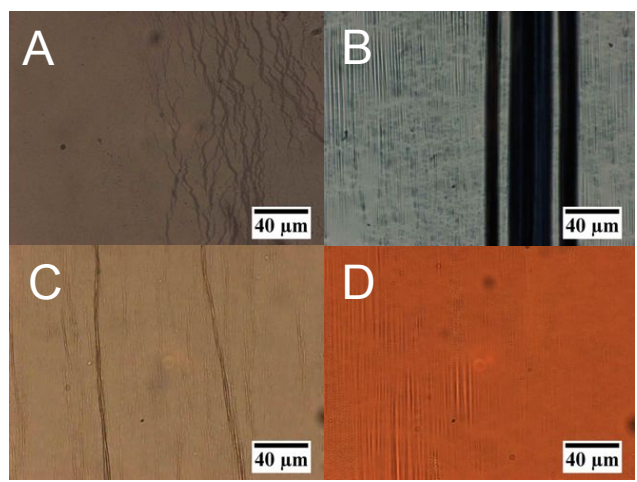

**Figure 9.** Optical images of nanoparticle film collapse for (**A**) 6 nm Au-dodecanethiol particles undergoing a monolayer to trilayer transition, (**B**) 6 nm Au-dodecanethiol nanoparticle trilayers buckling and folding, (**C**) 15 nm $Fe_3O_4$-oelic acid nanoparticle film folding and (**D**) 3 nm Ag-dodecanethiol nanoparticle film buckling.

The Au-nanoparticle film exhibits significantly higher compressive and shear moduli than those of $Fe_3O_4$ nanoparticle films. This may arise from the greater interdigitation of capping ligands on the Au-nanoparticles (dodecanthiol) compared to the ligands (oleic acid) on the $Fe_3O_4$ nanoparticles. Previous X-ray diffraction and TEM studies have determined the particle-particle separation of ~5nm Au-dodecanthiol nanoparticle films to be approximately the length of one ligand (1.8 nm), suggesting full interpenetration of the capping ligand shell.[3, 12] In contrast, for the larger $Fe_3O_4$ 15nm particles, this distance was found to be ~2.8 nm,[5] which is greater than the length of one oleic acid ligand (~2.0 nm) but less than two ligand lengths, suggesting that interdigitation occurs to a lower degree. Qualitatively, the oleic acid ligand may exhibit less interdigitation as it has a cis-double bond that reduces the ability of the ligands to interpenetrate. Ligand interpenetration increases the shear modulus by inhibiting the nanoparticles from moving past one another. Additionally the compressive moduli may be increased by denser packing of the interdigitated ligands, which prevents the ligands from effectively sliding against one another.

Alternatively, it is possible that the $Fe_3O_4$ particles have a lower ligand density, which decreases the degree of interactions between the ligands and reduces the stiffness of the film. This has been previously observed in freestanding nanoparticle membranes,[12] where it was shown that ligands bound more weakly to the $Fe_3O_4$ particles. Additionally, it is possible that the size of the nanoparticle also plays a role in the degree of attainable interdigitation through changing the curvature of nanoparticle surface, which in turn affects the conical volume around each individual ligand. These effects cannot be fully decoupled in this study.

Despite both being ligated with dodecanethiol, Au-nanoparticle films show significantly higher compressive and shear moduli than Ag-nanoparticle films. TEM measurements (Fig. 1) and previous X-ray studies[4] showed that this sample of Ag-nanoparticles displayed substantial polydispersity (~40 %), resulting in an absence of the hexagonal ordering observed in X-ray scattering measurements of the Au-nanoparticle samples.[3] We argue that the lower stiffness of the Ag-nanoparticle film arises from the wider size distribution of its constituent particles rather than differences in core materials.

This effect is clearly seen by comparing the original Au-nanoparticle sample with one that has been size-selected. Both the compressive and shear moduli increase more than twofold from the original to the size-selected Au-nanoparticle sample. This change arises as the size selection method removes particles near the extremes of the size spectrum, particularly larger particles or particle aggregates (Fig. 2). The difference in the measured macroscopic moduli develops from changes to the underlying nanoscopic structure of the film induced by changing particle polydispersity. A higher polydispersity in the nanoparticle sample may cause additional defects in the domains of the nanoparticle films, such as a decreasing nanoparticle island domain size, greater numbers or larger grain boundaries, or even phase separation of larger and smaller nanoparticles leading to a heterogeneous material. Through size selection, the polydispersity of the sample is reduced, which can in turn reduce the number of these types of defects in the film and increase the compressive and shear moduli.

**Nanoparticle and Lipid film Comparison.** An analogy can be drawn between the tail groups of lipids and the capping ligand choice for nanoparticles with respect to their effect on the mechanical properties of Langmuir monolayers. In both cases the packing ability of the organic molecules can influence strongly the mechanical properties of the corresponding self-assembled membranes. If the ligands or the tails are unsaturated and able to pack more densely together, then higher compressive moduli are achievable.

The compressive moduli of nanoparticle films are on the same order of magnitude as those of lipid films in the solid phase (DPPC). These similar moduli suggest qualitatively similar mechanical responses to an applied stress for both nanoparticle and lipid monolayers, as observed in their folding behaviors. However, nanoparticle monolayers show noticeably greater shear moduli ( ~70 mN/m for size selected Au-nanoparticles, ~ 40 mN/m for Au-nanoparticles, ~10-15 mN/m for Iron Oxide particles, ~17 mN/m for Ag) than that of DPPC (~4 mN/m).

This higher shear modulus of nanoparticle films in comparison to their lipid counterparts most likely arises from the interdigitation of the nanoparticle capping ligands. To some extent the nanoparticle sheet can be thought as a series of hard spheres (inorganic cores) that are linked together by an elastic mesh formed from interdigitation of the capping ligands. In contrast, individual lipids in a lipid membrane have their tails oriented upright in the same direction, and these tails do not tangle together, or otherwise interact to increase the shear modulus of the assembled membrane.

**Conclusion.** We have determined the two-dimensional compressive and shear moduli of optically uniform, macroscopic nanoparticle and lipid films using a two-Wilhelmy-plate pressure sensor method. This method allows the comparison of the relative stiffness of Langmuir films of differing materials. While nanoparticle and lipid films show compressive moduli on the same order of magnitude, nanoparticle films exhibit substantially higher shear modulus, likely due to interpenetration of their capping ligands.

Polydispersity significantly decreases the stiffness of the nanoparticle films, likely via increasing the number of defects in the particle packing. In addition, different types of ligands may influence the moduli, insofar as it determines the packing structure of the organic phase that fills the interstices between the nanoparticles. However, it is difficult to decouple this phenomenon from other effects.

The measurement of the anisotropic stress response of a Langmuir monolayer allows for the determination of the mechanical properties of a wide range of thin films, and can quantify different elastic responses arising from the structure and composition of different Langmuir monolayers.

**Corresponding Author**

*email: lin@cars.uchicago.edu

**Acknowledgements.** We thank M. Zhang, B.D. Leahy, and D. Sabulsky for their experimental help and insightful discussions and Prof. S. A. Rice for the lending of his equipment and fruitful discussions. This work was supported by the University of Chicago MRSEC of the NSF (DMR-0820054), the NSF Inter-American Materials Collaboration: Chicago-Chile, the University of Chicago MTSP (NIGMS/MSNRSA 5T32GM07281), and ChemMatCARS (NSF/DOE, Grant No. CHE-0822838).

References:

1. Holten-Andersen, N.; Michael Henderson, J.; Walther, F. J.; Waring, A. J.; Ruchala, P.; Notter, R. H.; Lee, K. Y. C., KL$_4$ peptide induces reversible collapse structures on multiple length scales in model lung surfactant. *Biophysical journal* **2011,** 101, 2957-65.
2. Lee, K. Y. C., Collapse mechanisms of Langmuir monolayers. *Annual review of physical chemistry* **2008,** 59, 771-91.
3. Schultz, D. G.; Lin, X.-M.; Li, D.; Gebhardt, J.; Meron, M.; Viccaro, P. J.; Lin, B., Structure, wrinkling, and reversibility of Langmuir monolayers of gold nanoparticles. *The journal of physical chemistry. B* **2006,** 110, 24522-9.
4. Kim, K.; Leahy, B. D.; Dai, Y.; Shpyrko, O.; Soltau, J. S.; Pelton, M.; Meron, M.; Lin, B., Governing factors in stress response of nanoparticle films on water surface. *Journal of Applied Physics* **2011,** 110, 102218.
5. Lee, D. K.; Kim, Y. H.; Kim, C. W.; Cha, H. G.; Kang, Y. S., Vast magnetic monolayer film with surfactant-stabilized Fe3O4 nanoparticles using Langmuir-Blodgett technique. *The journal of physical chemistry. B* **2007,** 111, 9288-93.
6. Vegso, K.; Siffalovic, P.; Majkova, E.; Jergel, M.; Benkovicova, M.; Kocsis, T.; Weis, M.; Luby, S.; Nygård, K.; Konovalov, O., Nonequilibrium phases of nanoparticle Langmuir films. *Langmuir : the ACS journal of surfaces and colloids* **2012,** 28, 10409-14.
7. Cheng, W.; Campolongo, M. J.; Cha, J. J.; Tan, S. J.; Umbach, C. C.; Muller, D. a.; Luo, D., Free-standing nanoparticle superlattice sheets controlled by DNA. *Nature materials* **2009,** 8, 519-25.
8. Leahy, B.; Pocivavsek, L.; Meron, M.; Lam, K.; Salas, D.; Viccaro, P.; Lee, K.; Lin, B., Geometric Stability and Elastic Response of a Supported Nanoparticle Film. *Physical Review Letters* **2010,** 105, 058301.
9. Tao, A.; Sinsermsuksakul, P.; Yang, P., Tunable plasmonic lattices of silver nanocrystals. *Nature nanotechnology* **2007,** 2, 435-40.
10. Dong, A.; Chen, J.; Vora, P. M.; Kikkawa, J. M.; Murray, C. B., Binary nanocrystal superlattice membranes self-assembled at the liquid-air interface. *Nature* **2010,** 466, 474-7.
11. Liao, J.; Zhou, Y.; Huang, C.; Wang, Y.; Peng, L., Fabrication, Transfer, and Transport Properties of Monolayered Freestanding Nanoparticle Sheets. *Small* **2011**, 583-587.

12. He, J.; Kanjanaboos, P.; Frazer, N. L.; Weis, A.; Lin, X.-M.; Jaeger, H. M., Fabrication and mechanical properties of large-scale freestanding nanoparticle membranes. *Small (Weinheim an der Bergstrasse, Germany)* **2010,** 6, 1449-56.
13. Kanjanaboos, P.; Joshi-imre, A.; Lin, X.-m.; Jaeger, H. M., Strain Patterning and Direct Measurement of Poisson's Ratio in Nanoparticle Monolayer Sheets. **2011**, 2567-2571.
14. Kanjanaboos, P.; Lin, X.-M.; Sader, J. E.; Rupich, S. M.; Jaeger, H. M.; Guest, J. R., Self-assembled nanoparticle drumhead resonators. *Nano letters* **2013,** 13, 2158-62.
15. Brus, L. E., Electron–electron and electron-hole interactions in small semiconductor crystallites: The size dependence of the lowest excited electronic state. *The Journal of Chemical Physics* **1984,** 80, 4403.
16. Anker, J.; Hall, W.; Lyandres, O.; Shah, N., Biosensing with plasmonic nanosensors. *Nature materials* **2008,** 7, 8-10.
17. Desireddy, A.; Joshi, C. P.; Sestak, M.; Little, S.; Kumar, S.; Podraza, N. J.; Marsillac, S.; Collins, R. W.; Bigioni, T. P., Wafer-scale self-assembled plasmonic thin films. *Thin Solid Films* **2011,** 519, 6077-6084.
18. Jain, P. K.; Lee, K. S.; El-Sayed, I. H.; El-Sayed, M. a., Calculated absorption and scattering properties of gold nanoparticles of different size, shape, and composition: applications in biological imaging and biomedicine. *The journal of physical chemistry. B* **2006,** 110, 7238-48.
19. Willets, K. a.; Van Duyne, R. P., Localized surface plasmon resonance spectroscopy and sensing. *Annual review of physical chemistry* **2007,** 58, 267-97.
20. Mikhaylova, M.; Kim, D. K.; Bobrysheva, N.; Osmolowsky, M.; Semenov, V.; Tsakalakos, T.; Muhammed, M., Superparamagnetism of magnetite nanoparticles: dependence on surface modification. *Langmuir : the ACS journal of surfaces and colloids* **2004,** 20, 2472-7.
21. Mukherji, S., Emerging use of nanostructure films containing capped gold nanoparticles in biosensors. *Nanotechnology, Science and Applications* **2010**, 171.
22. He, J.; Lin, X.-M.; Chan, H.; Vuković, L.; Král, P.; Jaeger, H. M., Diffusion and filtration properties of self-assembled gold nanocrystal membranes. *Nano letters* **2011,** 11, 2430-5.
23. Han, S.-T.; Zhou, Y.; Xu, Z.-X.; Huang, L.-B.; Yang, X.-B.; Roy, V. a. L., Microcontact printing of ultrahigh density gold nanoparticle monolayer for flexible flash memories. *Advanced materials (Deerfield Beach, Fla.)* **2012,** 24, 3556-61.
24. Paul, S.; Pearson, C.; Molloy, a.; Cousins, M. a.; Green, M.; Kolliopoulou, S.; Dimitrakis, P.; Normand, P.; Tsoukalas, D.; Petty, M. C., Langmuir–Blodgett Film Deposition of Metallic Nanoparticles and Their Application to Electronic Memory Structures. *Nano Letters* **2003,** 3, 533-536.
25. Boettcher, S. W.; Strandwitz, N. C.; Schierhorn, M.; Lock, N.; Lonergan, M. C.; Stucky, G. D., Tunable electronic interfaces between bulk semiconductors and ligand-stabilized nanoparticle assemblies. *Nature materials* **2007,** 6, 592-6.
26. Danov, K. D.; Kralchevsky, P. a.; Stoyanov, S. D., Elastic Langmuir layers and membranes subjected to unidirectional compression: wrinkling and collapse. *Langmuir : the ACS journal of surfaces and colloids* **2010,** 26, 143-55.
27. Pocivavsek, L.; Frey, S. L.; Krishan, K.; Gavrilov, K.; Ruchala, P.; Waring, A. J.; Walther, F. J.; Dennin, M.; Witten, T. a.; Lee, K. Y. C., Lateral stress relaxation and collapse in lipid monolayers. *Soft matter* **2008,** 4, 2019-2029.
28. Kaganer, V.; Möhwald, H.; Dutta, P., Structure and phase transitions in Langmuir monolayers. *Reviews of Modern Physics* **1999,** 71, 779-819.

29.	Khan, S. J.; Pierce, F.; Sorensen, C. M.; Chakrabarti, a., Self-assembly of ligated gold nanoparticles: phenomenological modeling and computer simulations. *Langmuir : the ACS journal of surfaces and colloids* **2009,** 25, 13861-8.
30.	Böker, A.; He, J.; Emrick, T.; Russell, T., Self-Assembly of Nanoparticles at Interfaces. *Soft Matter* **2007,** 3.
31.	Lin, Y.; Böker, A.; Skaff, H.; Cookson, D.; Dinsmore, a. D.; Emrick, T.; Russell, T. P., Nanoparticle assembly at fluid interfaces: structure and dynamics. *Langmuir : the ACS journal of surfaces and colloids* **2005,** 21, 191-4.
32.	Kulkarni, G. U.; Thomas, P. J.; Rao, C. N. R., Mesoscale organization of metal nanocrystals. *Pure and Applied Chemistry* **2002,** 74, 1581-1591.
33.	Petkov, J. T.; Gurkov, T. D., Dilatational and Shear Elasticity of Gel-like Protein Layers on Air / Water Interface. **2000,** 41, 3703-3711.
34.	Zang, D. Y.; Rio, E.; Langevin, D.; Wei, B.; Binks, B. P., Viscoelastic properties of silica nanoparticle monolayers at the air-water interface. *The European physical journal. E, Soft matter* **2010,** 31, 125-34.
35.	Weis, M.; Kopáni, M., Influence of vitamin C on alcohol binding to phospholipid monolayers. *European biophysics journal : EBJ* **2008,** 37, 893-901.
36.	Duncan, S. L.; Larson, R. G., Comparing experimental and simulated pressure-area isotherms for DPPC. *Biophysical journal* **2008,** 94, 2965-86.
37.	Espinosa, G.; López-Montero, I.; Monroy, F.; Langevin, D., Shear rheology of lipid monolayers and insights on membrane fluidity. *Proceedings of the National Academy of Sciences of the United States of America* **2011,** 108, 6008-6013.

**Table of Contents Only, TOC Figure**

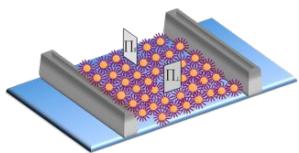
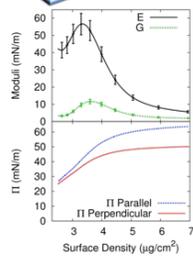